\documentclass[twocolumn,secnumarabic,superscriptaddress,amssymb, nobibnotes, aps, prl]{revtex4}

\usepackage{graphicx}
\usepackage{epsfig}

\begin{document}

\title{Coupling of Nitrogen-Vacancy Centers to Photonic Crystal Cavities in Monocrystalline Diamond}%

\author{Andrei Faraon}
\email{andrei.faraon@hp.com}
\affiliation{Hewlett Packard Laboratories, 1501 Page Mill Rd., Palo Alto, CA, 94304, USA}
\affiliation{California Institute of Technology, Department of Applied Physics and Material Science, 1200 East California Blvd., Pasadena, CA, 91125, USA}

\author{Charles Santori}
\affiliation{Hewlett Packard Laboratories, 1501 Page Mill Rd., Palo Alto, CA, 94304, USA}

\author{Zhihong Huang}
\affiliation{Hewlett Packard Laboratories, 1501 Page Mill Rd., Palo Alto, CA, 94304, USA}

\author{Victor M. Acosta}
\affiliation{Hewlett Packard Laboratories, 1501 Page Mill Rd., Palo Alto, CA, 94304, USA}

\author{Raymond G. Beausoleil}
\affiliation{Hewlett Packard Laboratories, 1501 Page Mill Rd., Palo Alto, CA, 94304, USA}

\begin{abstract}

The zero-phonon transition rate of a nitrogen-vacancy center is enhanced by a factor of $\sim 70$ by coupling to a photonic crystal resonator fabricated in monocrystalline diamond using standard semiconductor fabrication techniques. Photon correlation measurements on the spectrally filtered zero-phonon line show antibunching, a signature that the collected photoluminescence is emitted primarily by a single nitrogen-vacancy center. The linewidth of the coupled nitrogen-vacancy center and the spectral diffusion are characterized using high-resolution photoluminescence and photoluminescence excitation spectroscopy.

\end{abstract}

\maketitle


%

Nitrogen-vacancy (NV) centers in diamond represent one of the best test-beds for future quantum photonic technologies~\cite{ref:aharonovich2011dp,ref:obrien2009pqt} due to their outstanding properties as spin qubits~\cite{ref:balasubramanian2009usc} that can be coupled to optical photons~\cite{ref:togan2010qeb}. Applications include quantum information~\cite{ref:childress2005ftq} and electro-magnetic field sensing~\cite{ref:degen2008smf,ref:taylor2008hsd,ref:dolde2011esu}. For quantum information applications, multiple NV centers need to be interconnected. One approach relies on integrating NVs in photonic networks, where large-scale entanglement between NV centers is created using interference of identical photons at the zero-phonon line (ZPL) frequency. However, in bulk diamond the ZPL transition accounts for only a few percent of the NV emission. Here we report $\sim 70$ fold enhancement of the transition rate through the ZPL for a single NV coupled to a photonic crystal cavity in monocrystalline diamond. These cavities are fabricated using scalable semiconductor fabrication techniques and can be further coupled into photonic networks~\cite{ref:obrien2009pqt}.

\begin{figure}
\includegraphics[width=3in]{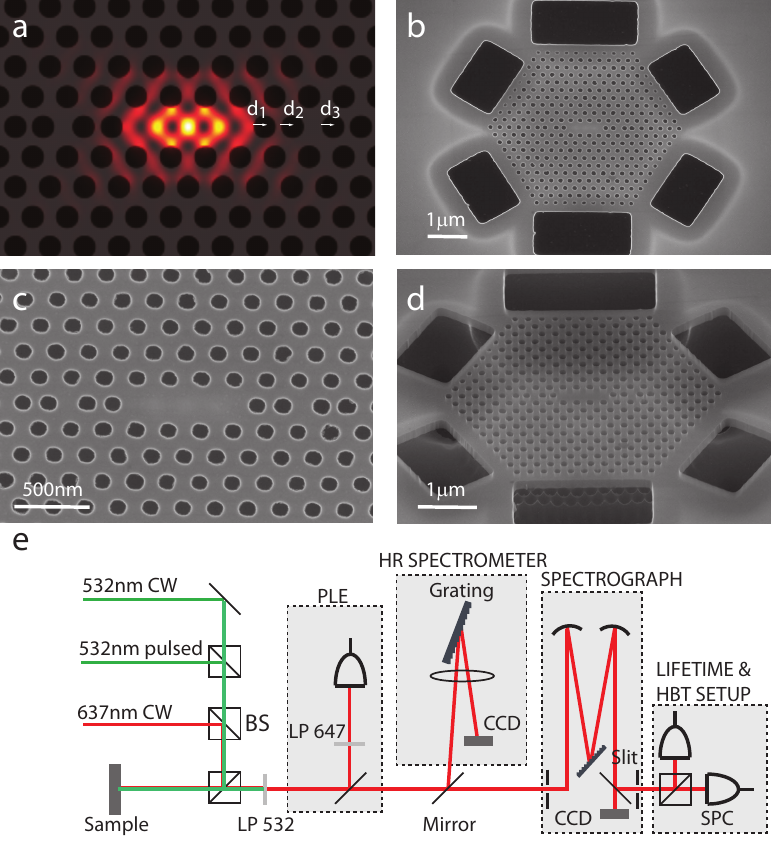}
\caption{(a) Color plot showing the electric field energy density in the fundamental mode of a linear three hole defect cavity. The lateral three holes are shifted laterally by $d_{1}$, $d_{2}$ and $d_{3}$ to increase the quality factor. (b-d) Scanning electron microscope images of a photonic crystal cavity fabricated in monocrystalline diamond. The black rectangles in (b) are openings etched in the membrane to facilitate etching underneath the membrane. The darker gray region around the openings indicates the extent of the etched Si under the diamond. (e) Schematic of the experimental setup. BS=beam splitter, LP=long pass filter, SPC=single photon counter, HBT=Hanbury-Brown and Twiss setup, HR=high resolution.}
\label{Fig:device_setup}
\end{figure}

The spontaneous emission rate enhancement of a particular dipole transition $i$ of an emitter coupled to a microresonator relative to the uniform dielectric medium of the resonator is enhanced by the factor $\left( \frac{\tau_{0}}{\tau_{\mathrm{leak}}} \right)_i + F_i$~\cite{ref:purcell1946sep}, where $1/\tau_{0}$ is the emission rate in the uniform dielectric medium, $1/\tau_{\mathrm{leak}}$ is the emission rate outside the cavity mode, and
\begin{equation}
F_i=F_{\mathrm{cav}}\left|\frac{\vec{E}(\vec{r}_i)\cdot\vec{\mu_i}}{\left|\vec{E}_{\mathrm{max}}\right|\left|\vec{\mu_i}\right|}\right|^2 \frac{1}{1+4Q^2(\frac{\lambda_i}{\lambda_{\mathrm{cav}}}-1)^2},
\label{eq:purcell}
\end{equation}
\noindent where $\vec{\mu_i}$ is the dipole moment, $\vec{E}(\vec{r}_i)$ is the local electric field at the emitter location $\vec{r}_i$, $\lambda_{\mathrm{cav}}$ is the cavity wavelength, $\lambda_{\mathrm{i}}$ is the emitter wavelength, and $\left|\vec{E}_{\mathrm{max}}\right|$ is the maximum value of the electric field in the resonator. For the case where the dipole is resonant with the cavity and also ideally positioned and oriented with respect to the local electric field, $F_i=F_{\mathrm{cav}}$ where,

\begin{equation}
F_{\mathrm{cav}}=\frac{3}{4 \pi ^2} \left(\frac{\lambda_{\mathrm{cav}}}{n}\right)^3\frac{Q}{V_{\mathrm{mode}}},
\label{eq:purcell_cav}
\end{equation}

\noindent $n$ is the refractive index and $V_{\mathrm{mode}}=\left( \int_{V}\epsilon(\vec{r}) \left| \vec{E} (\vec{r}) \right| ^2 d^3 \vec{r} \right) / \mathrm{max} \left(\epsilon(\vec{r}) \left| \vec{E} (\vec{r}) \right| ^2 \right) $ is the optical mode volume of the resonator, with $\epsilon(\vec{r})$ the electric permittivity at position $\vec{r}$. 

From Eq.~\ref{eq:purcell_cav} it can be inferred that the highest enhancement is achieved using optical resonators with small mode volume and high quality factors. Several groups are currently pursuing microcavities of various geometries for this purpose (see~\cite{ref:aharonovich2011dp} for a review). We previously reported coupling of the NV ZPL to the modes of a diamond microring resonator~\cite{ref:faraon2011rez}.  Compared to microrings, photonic crystal cavities provide confinement in a smaller mode volume and thus higher coupling rates. They can have mode volumes smaller than one cubic wavelength and the simulated quality factors can reach $Q=1.3 \times 10^6$~\cite{ref:bayn2008uhq,ref:tomljenovichanic2006dbp}. Photonic crystal cavities with quality factors up to 700 have been fabricated previously in single-crystal diamond and were coupled to silicon-vacancy colour centers~\cite{ref:riedrich-moller2011oat,ref:babinec2010dfi,ref:bayn2011ppc}. Hybrid systems involving photonic crystals fabricated in gallium phosphide have also been investigated for coupling to NV centers in diamond nano-crystals~\cite{ref:englund2010dcs,ref:wolters2010ezp}, but these NVs usually suffer from poor spectral properties.

A scanning electron microscope (SEM) image of a photonic crystal resonator similar to that used in the experiment is shown in Fig.~\ref{Fig:device_setup}(b-d) (the actual device is not shown because SEM imaging can affect the charge state of the NV). The device was designed using the parameters reported in Ref.~\cite{ref:tomljenovichanic2006dbp}, and consists of a linear three-hole defect photonic crystal cavity fabricated in a triangular lattice with period $a=218\, \mathrm{nm}$, hole radius $r = 0.29 a \sim 63\, \mathrm{nm}$ and membrane thickness $h=0.91a \sim (198\, \mathrm{nm})$ (the device shown in Fig.~\ref{Fig:device_setup}(b-d) has a photonic crystal with hole radius of $57 \, \mathrm{nm}$, smaller than the actual device). The three lateral holes are shifted outward by $d_{1}/a=0.21$, $d_{2}/a=0.02$ and $d_{3}/a=0.20$ (see Fig.~\ref{Fig:device_setup}(a)). Quality factors as high as $Q=6000$ can be achieved with this design, while the mode volume is $V_{\mathrm{mode}}\sim 0.88 (\lambda_{\mathrm{cav}}/n)^3 $ ($\lambda_{\mathrm{cav}}=637\, \mathrm{nm}$, $n=2.4$ for diamond). The simulated profile of the electric field energy density of the fundamental cavity mode is shown in Fig.~\ref{Fig:device_setup}(a). 

The fabrication recipe is based only on standard semiconductor micro-fabrication techniques, and is similar to the one reported in Ref.~\cite{ref:faraon2011rez} for microring resonators in single crystal diamond. A $\sim 200 \, \mathrm{nm}$ thick diamond membrane was obtained by thinning a $5\,\mu \mathrm{m}$ thick type IIa single-crystal diamond membrane (Element Six, $< 1 \, \mathrm{ppm}$ nitrogen) using reactive ion etching (RIE) in an oxygen plasma. Prior to the etching process the membrane was mounted on a Si substrate. After the membrane preparation, a silicon nitride ($100\,\mathrm{nm}$) layer was deposited on top and the photonic crystal device was patterned in this layer using electron-beam lithography and RIE. Two more etching steps were used to transfer the pattern from silicon nitride to diamond and then remove the excess silicon nitride. Finally, an isotropic dry etch removed the Si under the photonic crystal, and the membrane at the device location was left suspended (see Fig.~\ref{Fig:device_setup}(b-d)).

\begin{figure}
\centering
\includegraphics[width=3in]{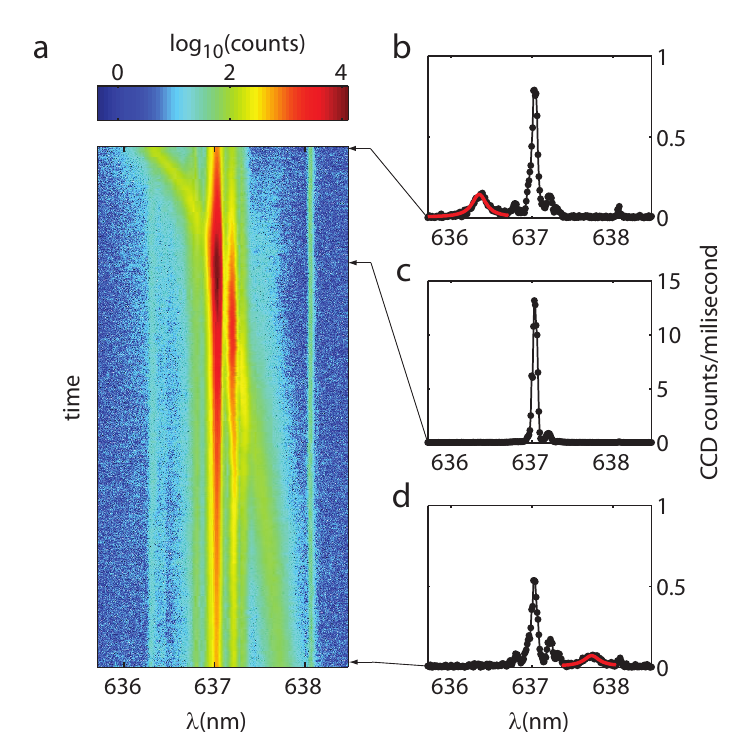}
\caption{(a) Color plot showing the photoluminescence spectra obtained while tuning the photonic crystal cavity over the ZPL of nitrogen vacancy centers located in the cavity. (b-d) Cross sections of the color plot. The cavity resonance is marked with the red Lorentzian fit.}
\label{Fig:PL}
\end{figure}

The chip was placed in a continuous flow liquid helium cryostat (Janis UHV ST-500) and was characterized using a confocal microscope setup [Fig.~\ref{Fig:device_setup}(e)]. The low temperature ($T<10K$) photoluminescence (PL) spectra under continuous wave (CW) green (532nm) excitation are shown in Fig.~\ref{Fig:PL}. The ZPLs from a few NV centers are visible at 637nm, together with the resonance of the cavity that is marked by the red Lorentzian fit. We estimated a quality factor of $Q \sim 3000$, about half of the the value predicted by simulations~\cite{ref:tomljenovichanic2006dbp}. The discrepancy is caused by imperfections in the fabrication process that causes non-uniformity in the hole radius and rugged sidewalls that are visible in Fig.~\ref{Fig:device_setup}(c). To observe coupling of the NVs to the cavity mode, the cavity is tuned into resonance with the ZPL using a technique where Xe gas is injected in the cryostat~\cite{ref:mosor2005spc}. Strong enhancement of the ZPL photoluminescence is observed (Fig.~\ref{Fig:PL}(c)) when the cavity becomes resonant with the ZPL of one of the NV
centers, a signature of the spontaneous emission rate enhancement.

\begin{figure}
\centering
\includegraphics[width=3in]{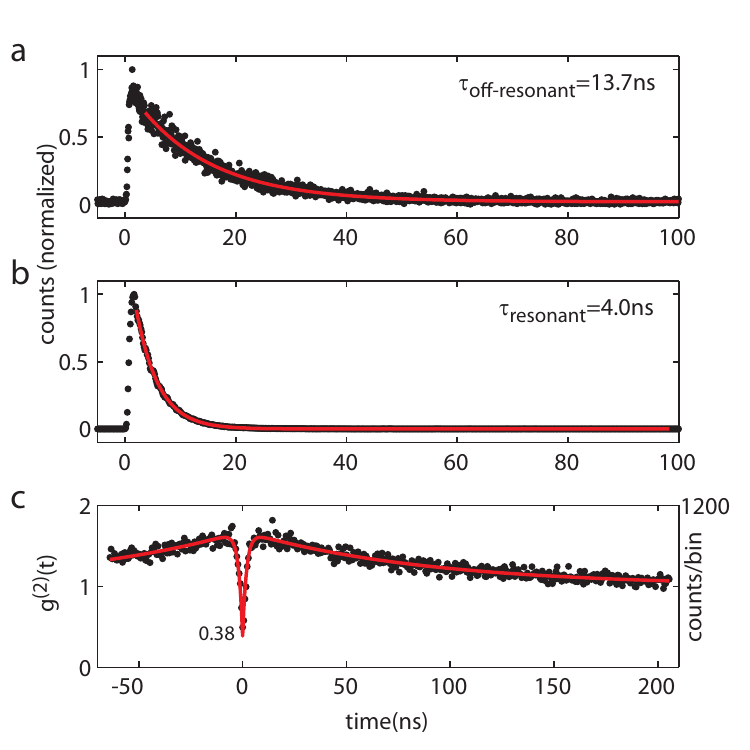}
\caption{(a) Measurement of the lifetime ($\tau_{\mathrm{0}}=13.7\, \mathrm{ns}$) of the NV center when not on resonance with the cavity. The measurement is performed on photoluminescence emitted in the ZPL. (b) Lifetime measurement when the NV is coupled to the cavity ($\tau_{\mathrm{c}}=4.0\, \mathrm{ns}$) (c) Second-order correlation ($g^{2}(\tau)$) measured on the zero-phonon line of the coupled NV. $g^{2}(0)=0.38$ is estimated from the fit to the data (red line) that takes into account multiple decay paths. The bin size in the histogram is $1.1\, \mathrm{ns}$.}
\label{Fig:lifetime_g2}
\end{figure}

Lifetime measurements were performed to determine more accurately the coupling strength between the cavity field and the ZPL transition. A pulsed green laser  (PicoQuant LDH-P-FA-530B, 532nm, 100ps pulse width, $\mathrm{100}\, \mu\mathrm{W}$ average power) was used to excite the NV center, and the PL emitted by the enhanced NV was filtered using a monochromator ($20\, \mathrm{GHz}$ bandwidth)  and sent to a photon counter (Perkin-Elmer SPCM). The lifetime is observed to change from $\tau_{\mathrm{off-resonant}}=13.7 \textrm{ns}$ in the off-resonant case (Fig.\ref{Fig:lifetime_g2}(a)) to $\tau_{\mathrm{resonant}}=4.0 \textrm{ns}$ when the ZPL is on resonance with the cavity (Fig.\ref{Fig:lifetime_g2}(b)). The off-resonant lifetime is slightly larger than the usual value of $\tau_{0}=12\, \mathrm{ns}$ generally observed in bulk diamond most likely due to the presence of the photonic bandgap that decreases the optical density of states\cite{ref:englund2005cse}. The zero-phonon spontaneous emission rate enhancement with respect to bulk can be derived using $F=(\tau _{0}/\tau _{\mathrm{resonant}}-\tau _{0}/\tau _{\mathrm{off-resonant}})/\xi_{\mathrm{ZPL}}$, where $\xi_{\mathrm{ZPL}}$ is the zero-phonon spontaneous emission fraction in bulk diamond~\cite{ref:faraon2011rez}. Considering that in bulk diamond $\xi_{\mathrm{ZPL}}\sim 3\%$ of the light is emitted in the ZPL, the lifetime modification corresponds to a spontaneous emission rate enhancemet of $1+F\sim \mathrm{70}$. The NV emits 70\% of the photons in the ZPL. The measured enhancement is a factor of four lower than the maximum expected value of $\sim 250$ derived from Eq.~\ref{eq:purcell_cav} using the device parameters, which is likely due to suboptimal alignment of the dipole with the optical field.

Emission lines from multiple NV centers are visible in the photoluminescence spectra in Fig.\ref{Fig:PL}, but due to the relatively low density of NV centers, only very few (ideally one) are coupled to the mode of the cavity. One method to identify whether the PL comes from a single emitter is to perform a second-order correlation measurement using a Hanbury-Brown Twiss (HBT) setup that consists of a beam splitter and two single-photon counters positioned at the two output ports (see Fig.~\ref{Fig:device_setup}(e)). Single photons (as emitted by a single NV) incident on the beam splitter can be directed only towards one of the counters, so the two counters cannot click at the same time. This can be tested by recording a time interval histogram for successive clicks on different counters. 


In this way, second-order correlation measurements were performed on the photons emitted in the ZPL when an NV was on resonance with the cavity. A continuous-wave ($380\, \mathrm{\mu W}$, 532nm) laser was used for excitation, and the collected light was spectrally filtered using the same setup as for the lifetime measurements. A total count rate of 30000/second was obtained from the spectrally filtered ZPL. An anti-bunching dip in the coincidence rate was observed at zero time delay, as shown in Fig.~\ref{Fig:lifetime_g2}(c). The dip extends over a few nanoseconds, and sits within a positive (``bunched") peak that extends over a timescale of one hundred nanoseconds. The behavior is typical for NV centers \cite{ref:kurtsiefer2000sss} and is caused by long-lived``shelving" states that do not emit light at the detection wavelength. The data were fitted using a formula for an emitter with three levels - a ground state, an excited state and a shelving state~\cite{ref:kurtsiefer2000sss}:

\begin{equation}
g^{(2)}(\tau)=1+c_{1} e^{-\tau/\tau_{1}}+c_{2} e^{-\tau/\tau_{2}},
\label{eq:g2}
\end{equation}

\noindent where the coefficients $c_{1},c_{2},\tau_{1},\tau_{2}$ depend on the pumping rate and the decay rates of the channels. The zero-time-delay auto-correlation was determined as $g^{(2)}(0)=0.38$, with $\tau_{1}=1.88\, \mathrm{ns}$ and $\tau_{2}=89\, \mathrm{ns}$ ($\tau_{1}$ is significantly shorter than the measured $\tau_{\mathrm{resonant}}$ because the NV is driven close to saturation with a high pump rate). The value $g^{(2)}(0)=0.38$ indicates that more than $75 \%$ of the photoluminescence comes from a single NV center well coupled to the photonic crystal resonator.

\begin{figure}
\centering
\includegraphics[width=3in]{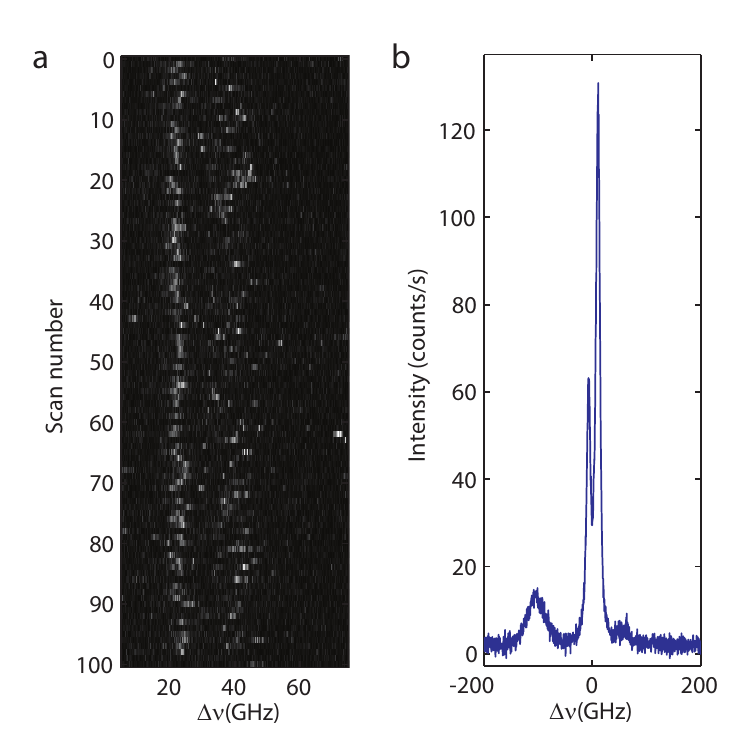}
\caption{(a) Photoluminescence excitation measurement at $T<10\, \mathrm{K}$ of the ZPL of the NV center coupled to the photonic crystal cavity.  (b) High-resolution photoluminescence measurement of the same NV.}
\label{Fig:ple_hrpl}
\end{figure}

Enhancing the NV ZPL emission as demonstrated above, is essential for increasing the success probability in quantum information schemes based on entangling NV centers via photon interference. An outstanding issue, however, is the spectral stability of NV centers which determines the entanglement fidelity in these schemes. We used high-resolution PL and photoluminescence excitation (PLE) spectroscopy to characterize the NV center coupled to the cavity. For the PLE measurements, a CW red laser diode was tuned to the NV ZPL transition for resonant excitation, and the PL emitted in the phonon sidebands was monitored while sweeping the laser frequency. The data are shown in Fig.\ref{Fig:ple_hrpl}(a), where two distinct spectral lines separated by $\sim 20 \, \mathrm{GHz}$ can be identified. The repeated scans, which are each followed by a 532 nm repump pulse, exhibit a combination of spectral diffusion behavior and a relatively broad width even from a single scan. The linewidths are approximately $4\, \mathrm{GHz}$ and $9\, \mathrm{GHz}$ for the left and right lines, or a factor of 100-225 greater than the cavity-modified lifetime limit of $40\, \mathrm{MHz}$. For comparison, the best-quality synthetic diamond with less than 5ppb nitrogen typically has spectral diffusion linewidths of ~100-200 MHz.

High-resolution PL spectroscopy was performed using a home-built setup, based on a large echelle grating, providing $0.8\, \mathrm{GHz}$ resolution. Fig.\ref{Fig:ple_hrpl}(a) shows a photoluminescence (PL) spectrum taken when the NV is enhanced by the cavity (the same conditions as in Fig.\ref{Fig:PL}(c)). As in the PLE measurement, two distinct peaks are observed. The FWHM of the main peak is approximately $8\, \mathrm{GHz}$, comparable to that observed in the PLE measurement. The presence of two peaks in the spectrum indicates either the presence of two NV centers, with one being significantly brighter than the other, or the strain-split (X/Y) branches of the same NV center. The broader peak on the left, presumably from another NV center, was outside of the filter passband used for the lifetime and photon-counting measurements described above.

In conclusion, we have demonstrated coupling of the zero-phonon line of a single NV center to a photonic crystal cavity fabricated in single-crystal diamond. The spontaneous emission into the ZPL is enhanced by a factor of $\sim 70$, which corresponds to $\sim 70 \%$ of the photons being emitted in the ZPL and into the cavity mode. The photonic crystal platfom allows for the possibility to dramatically improve these results with better designs ($Q > 10^{6}$ is theoretically possible~\cite{ref:bayn2008uhq}) and improved fabrication techniques. For example, with a mode volume of $\sim 1 (\lambda/n)^{3}$ and a quality factor of $\sim 10^{5}$, the system can operate at the onset of strong coupling regime, with more than $99 \%$ of the emission into the ZPL. Using high resolution spectroscopy we determined that the NV centers in this material have a single-scan linewidth of a few GHz and a spectral difusion under $532\, \mathrm{nm}$ of several GHz, which is not surprising for diamond of this quality (nitrogen concetration $ <1\, \mathrm{ppm}$). However, the measured linewidth is two orders of magnitude larger than the natural linewidth of NV centers, a serious issue for quantum information applications, where linewidths smaller than the ground state spin triplet splitting of $2.87\, \mathrm{GHz}$ are generally needed. Current and future effort is focused on fabricating similar devices in diamond material with much lower native nitrogen concentration ($< \, \mathrm{5 ppb}$) where the measured linewidth is much closer to the natural linewidth\cite{ref:fu2009odj}. If such spectral stability can be maintained in microfabricated structures, the resulting devices should enable efficient implementation of remote entanglement between NV centers via photon interference. In parallel, significant effort is dedicated towards integrating these cavities with electrodes used to stabilize the NV frequency against spectral diffusion~\cite{ref:acosta2011dso}, and towards the development of complex photonic networks as needed for quantum information devices for quantum chemistry~\cite{ref:lanyon2010tqc} and factoring~\cite{ref:nielsen2000qcq}.

This material is based upon work supported by the Defense Advanced Research Projects Agency under Award No. HR0011-09-1-0006 and The Regents of the University of California.


\end{document}